\documentclass[twocolumn,showpacs,preprintnumbers,amsmath,amssymb]{revtex4}

\usepackage{graphicx}
\usepackage{dcolumn}
\usepackage{bm}

\begin{document}

\preprint{}

\title{Dirac Equation on a Curved $2+1$ Dimensional Hypersurface}

\author{Mehmet Ali Olpak}
 \email{maliolpak@gmail.com}
\affiliation{Department of Physics, Middle East Technical University, 06531, Balgat, Ankara, Turkey}

\date{\today}

\begin{abstract}
Interest on $2+1$ dimensional electron systems has increased considerably after the realization of novel properties of graphene sheets, in which the behaviour of electrons is effectively described by relativistic equations. Having this fact in mind, the following problem is studied in this work: when a spin $1/2$ particle is constrained to move on a curved surface, is it possible to describe this particle without giving reference to the dimensions external to the surface? As a special case of this, a relativistic spin $1/2$ particle which is constrained to move on a $2+1$ dimensional hypersurface of the $3+1$ dimensional Minkowskian spacetime is considered, and an effective Dirac equation for this particle is derived using the so-called thin layer method. Some of the results are compared with those obtained in a previous work by M. Burgess and B. Jensen.   
\end{abstract}

\maketitle

\section{INTRODUCTION}
Graphene sheets, which are effectively $2+1$ dimensional objects, have some interesting properties that are relevant to both experimental studies and theoretical interest. The electrons in graphene exhibit a linear energy momentum dispersion as if they were massless Dirac particles \cite{Jackiw}. Some experimental studies on graphene has been awarded the 2010 Nobel Prize in Physics \cite{Geim}, and this further encourages both experimentalists and theorists in studying various aspects of effectively $2+1$ dimensional electron systems. One can also find field theoretic approaches to graphene in the literature \cite{Fialkovsky}. In this work, a $2+1$ dimensional model system with an electrostatic backgroud is considered and the dimensionality of the system is undertood as constraining a $3+1$ dimensional system onto a $2+1$ dimensional curved hypersurface by making use of the electrostatic background, as in \cite{Burgess}.  

In classical mechanics, there are certain types of problems in which geometrical constraints (which one may assume to be holonomic for present purposes) are imposed on physical systems. When dealing with such problems, the nature of interactions that generate those constraints are not generally taken into account directly, since the problems can be handled by other means (generally using the method of Lagrange multipliers). However, one does not deal with such cases in conventional problem sets of quantum mechanics, if, of course, one does not pay special attention to the issue. Beginning with Dirac, people tried to transpose that sort of problems to the framework of quantum mechanics (for example \cite{Dirac}, \cite{daCosta}, \cite{Ferrari}). Dirac developed a quantization procedure which involved handling the constraints within the commutation relations \cite{Dirac}, \cite{Ogawa}, \cite{Ikegami}. Other relevant works involved making explicit use of the geometrical relations arising from the existence of the constraints \cite{daCosta}, \cite{Ogawa}, \cite{Ferrari}, named as the \textquotedblleft thin layer" method by some authors \cite{Golovnev}.  However, the results obtained using those two approaches do not necessarily match with each other; and this is what one observes in the literature (see \cite{Ikegami}, \cite{Ogawatek}). There may be several possible solutions to this contradiction, which should eventually be verified experimentally. In any case, both approaches should be applied to a wider range of problems in order to be able to develop new ideas from the consequences.  

In this work, the above mentioned geometrical approach will be applied to Dirac equation in $3+1$ dimensional Minkowskian spacetime $M^{4}$ to reduce it to a $2+1$ equation, which will be viewed as if written on a curved hyper surface of $M^{4}$. The problem has been previously considered by M. Burgess and B. Jensen \cite{Burgess}, whose results are partly discussed in this work. The reason for choosing this approach is that it relies on the equation itself, without searching for a more general quantization procedure, and is thus only a limit of the $3+1$ dimensional equation. This makes it slightly more reliable in the sense that if it succeeds in the task given, then an appropriate (effective) Lagrangian formulation and a quantization procedure may be developed beginning from the resulting equation. However, the nature of possible interactions/mechanisms that may generate geometrical constraints are not studied in detail, and a simple elctrostatic background is assumed as in \cite{Burgess}. The reader will notice that results presented in this work do not exactly coincide with the counterparts given in \cite{Burgess}. 

\section{GEOMETRY}
Although the problem deserves a general treatment, we will choose a simple and specific geometry, which was also used by da Costa in \cite{daCosta} and Ferrari and Cuoghi in \cite{Ferrari}. We refer the reader to Mitchell's treatment \cite{Mitchell} which covers possibly the widest range of different geometries.

The above mentioned treatments \cite{daCosta, Ferrari} consider non-relativistic particles confined to move on a $2$ dimensional surface in $3$ dimensional Euclidean space, and the approach can be used to handle $N-1$ dimensional hypersurfaces of the $N$ dimensional Euclidean space directly. The geometry is expressed by the metric tensor transformed to a new basis via a general coordinate transformation. The curvilinear coordinates involved are expected to have no special properties, except that one of the coordinates is chosen to be orthogonal to the surface on which the particle will be constrained. Then, the relevant Schr\"{o}dinger equation is written in terms of these new coordinates, and the normal coordinate is squeezed so that one obtains an effective equation which involves only the surface coordinates. More explicitly \cite{tez}:
\begin{align}
& G_{\mu \nu} =\dfrac{\partial \mathbf{R}} {\partial q^{\mu}}\cdot \dfrac{\partial \mathbf{R}} {\partial q^{\nu}} \nonumber \\
& G_{i j} =\left(\dfrac{\partial \mathbf{r}} {\partial q^{i}}+q^{3}\dfrac{\partial \mathbf{N}} {\partial q^{i}} \right)\cdot \left(\dfrac{\partial \mathbf{r}} {\partial q^{j}}+q^{3}\dfrac{\partial \mathbf{N}} {\partial q^{j}} \right), \qquad i,j=1,2 \nonumber \\
& G_{i 3} =0, \qquad G_{3 3}=1.
\end{align}   
Here, $\mathbf{r}$ is the position vector of a generic point $P$ on the surface, $\mathbf{R}$ is the position vector of a point $Q$ just above the surface, $\mathbf{N}$ is the unit normal of the surface at $P$, $q^{\mu}$ are the curvilinear coordinates and the inner product is taken using the Euclidean metric $\delta _{ab}, \; a,b=1,2,3$. Notice that this expression is indeed exact with the coordinates chosen, but one may well write a Taylor expansion of position vectors \cite{tez}. However, this task is a little bit more complicated, and for the present purposes, the above expression is perfectly useful.  

In these coordinates, one obtains the following normalization integral for the wave function \cite{Ferrari, daCosta, tez}:
\begin{align}
& \int d^{3}x\Psi ^{*}(X)\Psi(X) =\int d^{3}q\sqrt{G}\Psi ^{*}(q)\Psi(q) \nonumber \\
&=\int d^{3}q\sqrt{g}\left (1+q^{3}Tr(\alpha)+(q^{3})^{2}det(\alpha)\right ) \Psi ^{*}(q)\Psi(q)=1,
\end{align}
where $X^{a}$ are the Cartesian coordinates, $G$ is the determinant of $G_{\mu \nu}$, $g$ is the determinant of the metric tensor $g_{ij}=\dfrac{\partial \mathbf{r}} {\partial q^{i}}\cdot \dfrac{\partial \mathbf{r}} {\partial q^{j}}$ induced on the surface, and $\alpha $ is the Weingarten matrix of the surface \cite{Ferrari} which is related to the extrinsic curvature of it and defined via \cite{daCosta, Ferrari, tez}:
\begin{align}
\dfrac{\partial \mathbf{N}}{\partial q^{i}}\equiv \alpha _{i}~^{j}\dfrac{\partial \mathbf{r}}{\partial q^{j}}.
\end{align}
Here, and in the following parts, Einstein summation convention is used. 

The redefinition of the wave function within the normalization integral is crucial, because the factor coming in front of the new wave function carries the signature of the \textquotedblleft external world" when substituted into the equation. With this procedure, after taking the limit $q^{3}\rightarrow 0$, one may calculate any observable without giving any reference to the external world \cite{tez}.

In the treatment for Dirac equation, this geometry will be assumed to correspond to the spatial part of the spacetime, and will be taken as constant in time, therefore causing no coupling between temporal and spatial parts. The only difference will be that Dirac equation is first order in all derivatives, while Schr\"{o}dinger equation is second order in spatial derivatives, so we will use the above objects up to first order in $q^{3}$, which also means there will survive no exact expressions for the metric tensor and related quantities. Such an expansion will do the job. 
 
\section{DIRAC EQUATION}
We consider an electron (or a spin half particle) in flat spacetime, but the use of general curvilinear coordinates requires writing the Dirac equation as if a curved spacetime is involved. This well known equation is written in curved spacetime in the following form \cite{Bertlmann}, \cite{tez}:
\begin{align}
(i\gamma ^{a}E_{a}~^{\mu}D_{\mu}-m)\psi =0,
\end{align}
where $\gamma ^{a}$ are the well known Dirac matrices satisfying:
\begin{align}
[\gamma ^{a},\gamma ^{b}]_{+}=2\eta ^{ab}
\end{align}
plus denoting the anti-commutator and $E_{a}~^{\mu}$ are known as the inverse vierbeins satisfying:
\begin{align}
G^{\mu \nu}=E_{a}~^{\mu}E_{b}~^{\nu}\eta ^{ab},\qquad \mu ,\; \nu ,\; a,\; b=0...3,
\end{align}
and $D_{\mu}$ is the appropriate covariant derivative which will explicitly be given below. Although the (inverse) vierbeins can be thought as coordinate transformation coefficients, this is not necessary in general. Indeed, these objects appear in the equation due to the requirement that Dirac equation should be written using an orthonormal basis of vectors or one-forms in order to be able to use the flat spacetime gamma matrices \cite{Bertlmann}. In addition, they are defined up to a local Lorentz transformation which leaves the relevant action invariant \cite{Bertlmann}.

In order to apply the thin layer method to this equation, one needs to expand the included objects in powers of $q^{3}$. This was performed before by Burgess and Jensen in \cite{Burgess}, but for surfaces having zero intrinsic curvature. When the surface has zero intrinsic curvature, one has the chance to use Cartesian coordinates in the vicinity of the surface \cite{Burgess}. However, this does not imply that the connection coefficients do not contribute to the geometric terms, contradicting with the statement of the authors. Indeed, making use of the algebra satisfied by the Dirac matrices, one observes that:
\begin{align}
\gamma ^{c}E_{c}~^{\mu}\omega _{ab\mu}[\gamma ^{a},\gamma ^{b}]=2\gamma ^{a}(\nabla _{\mu}E_{a}~^{\mu})
\end{align}
due to 
\begin{align}
\gamma ^{a}[\gamma ^{b},\gamma ^{c}]=[\gamma ^{c},\gamma ^{a}]\gamma ^{b}-2\eta ^{ac}\gamma ^{b}+2\eta ^{bc}\gamma ^{a}
\label{usefulrel}
\end{align}
regardless of dimension of the spacetime.  

Now, let us write the equation explicitly. First, it should be noted that, the vierbeins and inverses satisfy the following \cite{Bertlmann}:
\begin{align}
e^{a}~_{\mu}E_{a}~^{\nu}=\delta ^{\nu}~_{\mu}, \; e^{a}~_{\mu}E_{b}~^{\mu}=\delta ^{a}~_{b},
\end{align}
and the covariant derivative is given as \cite{Bertlmann}, \cite{tez}:
\begin{align}
& D_{\mu}=\partial _{\mu}+\omega _{\mu};\nonumber \\
& \omega _{\mu}=\dfrac{1}{8}\omega _{ab\mu}[\gamma ^{a},\gamma ^{b}];\nonumber \\
& \omega ^{a}~_{b\mu}=-E_{b}~^{\nu}\nabla _{\mu}e^{a}~_{\nu},
\end{align}
where $\nabla $ is the covariant derivative in the so called coordinate basis \cite{Bertlmann}, which is not necessarily orthonormal, in other words:
\begin{align}
\nabla _{\mu}V^{\nu}=\partial _{\mu}V^{\nu}+\Gamma ^{\nu}_{\mu \lambda}V^{\lambda}.
\end{align}

Now, we introduce the following set of vierbeins \cite{tez}(somehow different than in the reference):
\begin{align}
& e^{a}~_{i}=\partial _{i} x^{a}+q^{3}\partial _{i}n^{a}=\partial _{i} x^{a}+q^{3}\alpha _{i}~^{k}\partial _{k}x^{a}, \nonumber \\
& e^{0}~_{0}=1, \; e^{a}~_{3}=N^{a}, \; a=1,2,3,\; \mbox{others}=0,
\end{align}
whose inverses are given by (to first order in $q^{3}$):
\begin{align}
& E_{a}~^{i}=\tilde{E}_{a}~^{i}-q^{3}\alpha _{k}~^{i}\tilde{E}_{a}~^{k}, \; E_{0}~^{0}=1, \; E_{a}~^{3}=N_{a}, \nonumber \\ 
& a=1,2,3, \; \mbox{others}=0, 
\end{align}
where the over tilde implies that the quantity is evaluated on the surface and $x^{a}$ are the flat coordinates of the generic point $P$ lying on the surface. This notation implies: 
\begin{align}
\tilde{e}^{a}~_{\mu}\equiv (\partial _{\mu}x^{a})_{0}.
\end{align}
We also redefine the spinor in the following way \cite{Burgess}, \cite{tez}:
\begin{align}
\chi \equiv \psi \sqrt{1+q^{3}Tr(\alpha )}.
\end{align}
where the geometrical relations give $Tr(\alpha )=\tilde{\Gamma }^{i}_{i3}, \, i=1,2$. Using these objects, one obtains the following equation near $q^{3}=0$:
\begin{align}
& i\Bigg(\gamma ^{0}\partial _{0}\chi +\gamma ^{A}\tilde{E}_{A}~^{i}\partial _{i}\chi +\dfrac{1}{4}\gamma ^{A}(\tilde{\nabla}_{i}\tilde{E}_{A}~^{i})\chi \nonumber \\
& -\dfrac{1}{4}\gamma ^{A}\tilde{N}_{A}Tr(\alpha )\chi +\gamma ^{A}\tilde{N}_{A}\partial _{3}\chi \Bigg)-m\chi =0,\nonumber \\
& A=1,2,3; \; i=1,2,
\label{curved1}
\end{align}
where the over tildes imply the object is evaluated at $q^{3}=0$, and $x^{c}$ are the flat coordinates as introduced before. In order to obtain this equation, one makes use of (\ref{usefulrel}) and:
\begin{align}
N^{a}E_{a}~^{i}=0.
\end{align}

Here, the term $-\dfrac{i}{4}\gamma ^{A}\tilde{N}_{A}Tr(\alpha )\tilde{\chi}$ appears due to the existence of an external world, that is, it is the residue of the constrained dimension within the equation. there are two contributions to this term: one from the redefiniton of the spinor and the other from the connection coefficients. This term is the  analogue of the geometric potential which appears in the Schr\"{o}dinger equation in \cite{daCosta, Ferrari}; however, while that geometric potential was a scalar function, this seems like a vector potential, but one which does not have a temporal component. This seems weird, of course; in the case of a specified interaction, the meaning of this term may become more clear.  

Burgess and Jensen assume that their resulting equation is separable, and they consider two equations in \cite{Burgess}, one involving only the normal coordinate, and the other involving surface parameters. They conclude that the solutions behave like a shifted Gaussian along the normal direction. Though we proceeded in a somewhat different way, we may give an argument which is also in accordance with that of Burgess and Jensen; that is, we may assert that the solution to the $3+1$ dimensional equation in the same coordinates should have an extremum, more precisely a maximum at $q^{3}=0$, if it is really possible to constrain the particle to that surface. One may also take this argument in the following way: $\overline{\chi}(t,\mathbf{r})\chi (t,\mathbf{r})$ is the probability density for the particle to be found at time $t$ at the point $\mathbf{r}$ (remember that the spinor has been redefined so that the measure of the normalization integral involves only the determinant of the metric induced on the surface), and so, if the particle is constrained on a surface, then \textquotedblleft it is most probably on the surface\textquotedblright . So, the probability density should be maximum at $q^{3}=0$. These statements do not directly imply that $\chi $ has a maximum at $q^{3}=0$; however, it will be observed that a Gaussian distribution around the surface necessarily results in that. To see this, let us introduce an electrostatic potential as a background, and derive the consequences:
\begin{align}
& i\Bigg(\gamma ^{0}\partial _{0}+\gamma ^{A}\tilde{E}_{A}~^{i}\partial _{i}+\dfrac{1}{4}\gamma ^{A}(\tilde{\nabla}_{i}\tilde{E}_{A}~^{i})\nonumber \\
& -\dfrac{1}{4}\gamma ^{A}\tilde{N}_{A}Tr(\alpha )+\gamma ^{A}\tilde{N}_{A}\partial _{3}\Bigg)\chi \nonumber \\
& +\Big(e\gamma ^{0}A_{0}(q^{3})-m\Big)\chi =0, \; A=1,2,3; \; i=1,2,
\end{align}
where $e$ is the charge of the particle. Following the arguments of \cite{Burgess}, one can identify this electrostatic potential with some $\lambda $ times $q^{3}$, and expect that the behaviour of the spinor along the normal coordinate to be determined by the relevant derivative and the coupling term:
\begin{align}
(\gamma ^{0}e\lambda q^{3}-i\gamma ^{A}\tilde{N}_{A}\partial _{3})\chi =K\chi .
\label{normaleq}
\end{align}

Up to this point, every term but the normal dervative has been evaluated at $q^{3}=0$. However, there is an order $q^{3}$ term now. Here, one has to assume that $e\lambda q^{3}$ also has an $O(1)$ contribution, which is consistent with the physics of the problem \cite{Burgess}. But there is another issue here: components of the normal vector are functions of the surface coordinates, so how can one get rid of the dependence on these coordinates in this equation? One can simply expand the spinor in terms of the eigenvectors of the matrix $i\gamma ^{A}\tilde{N}_{A}$ and assume seperability of the equation for each term in the expansion. The eigenvalues of this matrix are $\pm 1$, and assuming a Gaussian-like dependence for all componentes of the spinor imply:
\begin{align}
\chi \equiv exp[-a(q^{3})^2+bq^{3}]\sum _{r=1}^{4}h_{r}(t,q^{i})V_{r}(q^{i}),
\end{align}
where $a>0$ and $V_{r}(t,q^{i})$ are the eigenvectors of $i\gamma ^{A}\tilde{N}_{A}$. If the equation is seperable, $K$ must be a constant matrix. It is a chance that one does not have to be that strict at all: $K$ can be a function of time and the surface coordinates only. Putting this expression into (\ref{normaleq}), one obtains:
\begin{align}
& exp[-a(q^{3})^2+bq^{3}]\sum _{r=1}^{4}\Big[e\lambda \gamma ^{0}\nonumber \\ 
& +c_{r}(-2aq^{3}+b)-k_{r}\Big]h_{r}(t,q^{i})V_{r}(q^{i})=0,
\end{align}
where $c_{r}$ and $k_{r}$ are the eigenvalues and corresponing seperation constants (or functions) respectively. Imposing that the probability distribution has a maximum at $q^{3}=0$ amounts to $b=0$, which also amounts to $k_{r}=0, \, \forall r$. This contradicts with the result in \cite{Burgess}, since in that reference the Gaussian distribution is shifted and the effect is interpreted as a contribution to the effective mass of the states after constraining the system. The solution is then a pure Gaussian with $a=-\dfrac{e\lambda }{2}$. The remaining (tangential) equation is then:
\begin{align}
& \sum _{r=1}^{4}\Bigg[i\Big(\gamma ^{0}\partial _{0}+\gamma ^{A}\tilde{E}_{A}~^{i}\partial _{i}+\dfrac{1}{4}\gamma ^{A}(\tilde{\nabla}_{i}\tilde{E}_{A}~^{i})\Big)\nonumber \\
& -\dfrac{c_{r}}{4}Tr(\alpha )-m\Bigg]\Big(h_{r}(t,q^{i})V_{r}(q^{i})\Big)=0.
\label{sontan}
\end{align}

The geometric term can still be interpreted as an effective contribution to mass \cite{Burgess}, and the equation can be written as a system of coupled $2\times 2$ equations for two 2-spinors. One expects that these equations should be decoupled for two linear combinations of those 2-spinors. Consider, for example, the following $4$ representation for the gamma matrices:
\begin{align}
\gamma ^{0}=\begin{bmatrix} 1 & 0 \\ 0 & -1 \end{bmatrix} , \; \gamma ^{A}=\begin{bmatrix} 0 & i\sigma ^{A} \\ -i\sigma ^{A} & 0  \end{bmatrix},
\end{align}
where all entries are $2\times 2$ matrices and $\sigma ^{A}, \, A=1,2,3$ are the well known Pauli matrices satisfying:
\begin{align}
[\sigma ^{A},\sigma ^{B}]_{+}=2\delta ^{AB}.
\end{align}
 
In this representation, after multiplying (\ref{sontan}) with $i\gamma ^{A}\tilde{N}_{A}$, the equations appear as the following:
\begin{align}
& i\Big(-c_{r}\sigma ^{A}\tilde{N}_{A}\partial _{0}+\sigma ^{A}\tilde{E}_{A}~^{i}\partial _{i}+\dfrac{1}{4}\sigma ^{A}(\tilde{\nabla}_{i}\tilde{E}_{A}~^{i})\Big)v_{r}(t,q^{i})\nonumber \\
& +\Big(\dfrac{c_{r}}{4}Tr(\alpha )+m\Big)u_{r}(t,q^{i})=0, \nonumber \\
& i\Big(c_{r}\sigma ^{A}\tilde{N}_{A}\partial _{0}+\sigma ^{A}\tilde{E}_{A}~^{i}\partial _{i}+\dfrac{1}{4}\sigma ^{A}(\tilde{\nabla}_{i}\tilde{E}_{A}~^{i})\Big)u_{r}(t,q^{i})\nonumber \\
& +\Big(\dfrac{c_{r}}{4}Tr(\alpha )+m\Big)v_{r}(t,q^{i})=0,
\end{align}
where $h_{r}(t,q^{i})V_{r}(q^{i})\equiv \begin{bmatrix}u_{r}(t,q^{i}) \\ v_{r}(t,q^{i})\end{bmatrix}$. Though the equations are still coupled, this representation in $3+1$ dimensions naturally suggest the follwing representation for the (coordinate dependent) gamma matrices in $2+1$ dimensions:
\begin{align}
\Gamma ^{0}=\pm \sigma ^{A}\tilde{N}_{A}, \, \Gamma ^{i}=\pm i\sigma ^{A}\tilde{E}_{A}~^{i}, \, A=1,2,3, \, i=1,2, 
\end{align}
which clearly satisfy
\begin{align}
[\Gamma ^{\alpha},\Gamma ^{\beta}]_{+}=2g^{\alpha \beta},
\end{align}
and $g^{\alpha \beta}$ is clearly the induced metric on the $2+1$ dimensional hypersurface.

Here, there is one interesting fact: the charge of the particle does not appear in the reduced equations! This is interesting, because it suggests the idea that the electrostatic interaction in $3+1$ dimensions seems to mimic gravity in $2+1$ dimensions, since the considered geometry is curved and the extrinsic contributions can be absorbed into the mass term. This fact then brings a new question: how can one incorporate gravitation effectively in this picture? For sure, the answer will involve the complete set of conditions on the nature of the solutions to the tangential equation.
   
However, in the case of a magnetic interaction, one may not be able to reduce the equation to $2+1$ dimensions. This is suggested even by non-relativistic spin-magnetic field interaction: 
\begin{align}
\widehat{H}_{I}=-\dfrac{e}{2mc}(\nabla \times \mathbf{A}\cdot \mathbf{S})_{0}
\end{align}
where $\mathbf{A}$ is the electromagnetic vector potential. The cross product counts the number of spacetime dimensions, and a possible reduction scheme will most probably be gauge dependent \cite{tez}. Beyond speculations, one has to check what happens in this case. 

\section{CONCLUSION}
In this work, the treatment of Burgess and Jensen has been generalized to surfaces having nonzero intrinsic curvature, and the arguments on the origin of the geometric term and the behaviour of the solutions along the normal direction have been corrected. However, there still are a number of unanswered questions for this case. First of all, solutions to the reduced equations should be derived and interpreted in a future work, and limitations on possible experimental realizations should be discussed. Secondly, one has to examine whether there are any geometrical constraints on the relevant surface, as done by Burgess and Jensen in their treatment. The last, and may be the most interesting unanswered question is, as indicated in the previous section, is the one concerning similarity with $2+1$ gravity.

One can think of more interesting but complicated scenarios in terms of the interaction constraining the system into $2+1$ dimensions. However, the treatments seem to be limited in the case of single particle equations. A more general, may be a field theoretic treatment can be studied in order to develop deeper, and experimentally verifiable ideas. 

\section{ACKNOWLEDGEMENTS}
The author is grateful to Bayram Tekin who suggested this problem and helped throughout its solution. Detailed derivations that led to the discussion in this work can be found in our M. Sc. Thesis \cite{tez}. The author has special thanks to Ozgur Sarioglu for indicating new research directions initiating from the problem. The author would also like to thank T. \c{C}a\u{g}r{\i} \c{S}i\c{s}man for his aid in formatting issues. 

\newpage 

\end{document}